\documentclass{article}
\usepackage{amsfonts}
\usepackage{amssymb}
\usepackage{amsmath}
\usepackage{graphicx}
\usepackage{epstopdf}
\usepackage{slashed}
\usepackage{setspace}

\newcommand{\reef}[1]{(\ref{#1})}

\newcommand{\M}{{1{\rm m}}}
\newcommand{\MMe}{{2{\rm me}}}
\newcommand{\MMh}{{2{\rm mh}}}
\newcommand{\MMM}{{3{\rm m}}}

\newcommand{\ca}{{\cal A}}

\newcommand{\cn}{{\cal N}}

\newcommand{\cR}{{\cal R}}
\newcommand{\co}{{\cal O}}
\newcommand{\cp}{{\cal P}}
\newcommand{\cs}{{\cal S}}

\newcommand{\be}{\begin{equation}}
\newcommand{\ee}{\end{equation}}

\newcommand{\PP}{{\mathbb{P}}}
\newcommand{\Ord}{O}
\newcommand{\Li}{\mathrm{Li}}

\def\be{\begin{equation}}
\def\ee{\end{equation}}
\def\bea{\begin{eqnarray}}
\def\eea{\end{eqnarray}}
\def\ba{\begin{array}}
\def\ea{\end{array}}
\def\bd{\begin{displaymath}}
\def\ed{\end{displaymath}}

\def\eg{{\it e.g.~}}
\def\ie{{\it i.e.~}}

\def\a{\alpha}
\def\b{\beta}

\def\d{\delta}
\def\e{\epsilon}           

\def\h{\eta}

\def\l{\lambda}
\def\m{\mu}

\def\D{\Delta}

\def\>{\rangle} 
\def\<{\langle} 
\def\Dsl{D \hskip-.6em \raise1pt\hbox{$ / $ } }
\def\to{\rightarrow}

\def\lab{\label}

\newcommand{\eps}{\epsilon}

\newcommand{\cyc}{{\rm cyc}}

\begin{document}

\setstretch{1.05}

\begin{titlepage}

\begin{flushright}
MIT-CTP-4042 \\
\end{flushright}
\vspace{1cm}

\begin{center}
{\Large\bf Dual conformal symmetry of 1-loop} \\[2.5mm]
{\Large\bf  NMHV amplitudes in $\cn =4$ SYM theory}  \\[2.5mm]
\vspace{1cm}
{\bf Henriette Elvang${}^{a}$,
Daniel Z.~Freedman${}^{b,c}$, Michael Kiermaier$^{b}$} \\
\vspace{0.7cm}
{{${}^{a}${\it School of Natural Sciences}\\
         {\it Institute for Advanced Study}\\
         {\it Princeton, NJ 08540, USA}}\\[5mm]
{${}^{b}${\it Center for Theoretical Physics}}\\
{${}^{c}${\it Department of Mathematics}}\\
         {\it Massachusetts Institute of Technology}\\
         {\it 77 Massachusetts Avenue}\\
         {\it Cambridge, MA 02139, USA}}\\[5mm]
{\small \tt  elvang@ias.edu,
 dzf@math.mit.edu, mkiermai@mit.edu}
\end{center}
\vskip .3truecm

\begin{abstract}
\noindent
We prove that 1-loop $n$-point NMHV superamplitudes in  $\cn =4$ SYM theory are dual conformal
covariant for all numbers $n$ of external particles (after regularization and subtraction of IR divergences).  This property was previously established for $n \le 9$ in arXiv:0808.0491.
We derive an explicit representation of these superamplitudes in terms of dual conformal cross-ratios.
We also show that all the 1-loop `box coefficients' obtained from maximal cuts of N$^k$MHV
$n$-point functions are covariant under dual
conformal transformations.
\end{abstract}
\end{titlepage}

\tableofcontents
\newpage

\setcounter{equation}{0}
\section{Introduction}
Dual conformal symmetry is a newly discovered symmetry of on-shell
scattering amplitudes in $\cn =4$ SYM theory.
We will survey the development
of the subject briefly below, but first we state the particular focus and result of this
technical note.

We focus on the proposed \cite{DHKS1}  dual conformal (and dual superconformal) symmetries
of color-ordered planar $n$-point superamplitudes   $\ca^{{\rm N}^k {\rm MHV}}_n.$
These encode the values of the $n$-gluon  amplitudes with $k+2$ negative and $n-k-2$ positive helicity
gluons, plus all amplitudes of the  $\cn =4$  theory related to these $n$-gluon amplitudes by conventional Poincar\'e supersymmetry.
 Each external particle is described  within the superamplitude by a null 4-momentum
$p_i^\m$, the associated Weyl spinors  $\l_{\a i},~\tilde{\l}^{\dot{\a}}_i$ and a Grassmann bookkeeping variable $\h^a_i$ with $i=1,\ldots n, ~ a= 1,2,3,4$.  We  use Dirac `kets' to denote
spinor variables, \eg  $\l_{\a i} \to  |i\>_\a$ and $\tilde{\l}^{\dot{\a}}_i \to |i]^{\dot{\a}}$.
Dual  superconformal transformations act most simply on the dual bosonic and fermionic `zone variables'
$x_i^\m$ and  
$|\theta^a_i\>$ defined by
\be \lab{zone}
p_i^\m \equiv x_i^\m - x_{i+1}^\m\,, ~\qquad |i\> \h_i^a \equiv |\theta^a_i\> - |\theta^a_{i+1}\>\,,
~\qquad i=1,\ldots, n\,.
\ee

It is a bit complicated to describe the action of this symmetry on loop amplitudes because these have infrared divergences, which break conformal properties.  The MHV superamplitude $\ca^{\rm MHV}_n$ is distinguished by its simplicity in the tree approximation and by the fact that,  in a sense which we describe momentarily, it captures the `universal'
external momentum dependence of the IR divergence of all
loop amplitudes.  For these reasons it is useful to discuss general N$^k$MHV processes in terms of   the ratio of superamplitudes:
\be \lab{ratio}
\ca^{{\rm N}^k {\rm MHV}}_n = \ca^{\rm MHV}_n ({\cal R}^{{\rm N}^k {\rm MHV}}_n+ \co(\e)),
\ee
with ${\cal R}^{\rm MHV}_n =1.$ We use $\ca^{{\rm N}^k {\rm MHV}}_{n,L}$, $\cR^{{\rm N}^k {\rm MHV}}_{n,L}$ to denote the $L$-loop contributions.
It is known that in tree approximation all $\ca^{{\rm N}^k {\rm MHV}}_{n,0}$ transform covariantly
under dual superconformal transformations \cite{DHKS1,QMC1} and that the tree-level ratio ${\cal R}^{{\rm N}^k {\rm MHV}}_{n,0}$ is dual superconformal invariant and can be expressed  \cite{DH}
in terms of a set of superconformal invariants.  These invariants were defined at the NMHV level in
 \cite{DHKS1}  and for general $k$ in \cite{DH}.

It is a consequence of the Ward identities of conventional Poincar\'e supersymmetry that
at 1-loop order (and beyond),  $\ca^{\rm MHV}_n$ is the product of the tree factor   $ \ca^{\rm MHV}_{n,0} $ times an IR divergent  function of commuting variables only, \ie no $\h^a_i$.  It is the IR finite quantity ${\cal R}^{{\rm N}^k {\rm MHV}}_n$  which apparently enjoys simple  properties under dual conformal symmetry. It was conjectured \cite{DHKS1,DHKS2} that ${\cal R}^{{\rm N}^k {\rm MHV}}_n$ is dual conformal invariant
to all orders and for all $k$.\footnote{Concerning dual {\em super}conformal invariance, see brief discussion in section~\ref{sec7}.}
At 1-loop order this conjecture was proven for NMHV amplitudes for $n=6,7,8$ or $9$ external particles.  Explicit representations in terms of  dual  invariant cross-ratio variables were given for $n=6,7.$
\emph{In this note we prove that the 1-loop ratio ${\cal R}^{{\rm  NMHV}}_{n,1}$ is  dual conformal invariant for all $n$
at 1-loop order, and  we obtain explicit expressions in terms of
 dual conformal cross-ratios.}

Our approach is a generalization to all $n$ of the methods developed in \cite{DHKS2}.
Initially, we obtain a representation of the ratio $ {\cal R}^{{\rm  NMHV}}_{n,1}$ that contains $n(n\!-\!2)(n\!-\!3)/2$ terms, namely one for each dual superconformal invariant $R_{rst}$. Each invariant is multiplied by a linear combination of scalar box integrals, which is neither IR finite nor dual conformal invariant! 
The invariants $R_{rst}$ are not independent, however, and we recursively solve the linear relationships among them  to eliminate the terms of $2n^2-10n-1$ invariants from the ratio. The final form we obtain is then both IR finite and dual conformal invariant, and we express it in terms of dual conformal cross-ratios.

Two strands of investigation have led to the present understanding of dual conformal symmetry.  At weak coupling, there were
 early hints of dual conformal  behavior in multi-loop diagrams  for the off-shell 4-point functions
analyzed in \cite{magic}.  Then the AdS/CFT correspondence was applied in \cite{AM1}
(see also \cite{AM2,Alects,AR,BM,Beisert}) to obtain the strong coupling limit of the planar 4-gluon amplitude.  The
amplitude was found to agree with the expectation value of the Wilson loop evaluated on a closed polygonal path with vertices at $x_i$, and with  $x_i -x_{i+1}= p_i$ identified with the null momenta of the gluons.   A polygonal Wilson loop embodies \cite{DHKS4} an anomalous form of conformal symmetry due to the cusps at the vertices. 
Dual conformal symmetry of the scattering amplitude is simply conventional conformal symmetry of the Wilson loop.

The second line of investigation
concerns dual conformal symmetry of on-shell amplitudes at weak coupling.  The symmetry was an important aspect of studies   \cite{DKS,Brandhuber:2007yx,DHKS3,DHKS4,DHKS5,BDKR,DHKS6,QMC2} of the relation between higher-loop MHV amplitudes and polygonal Wilson loops which were motivated by the BDS conjecture \cite{BDS}.  The extension to dual superconformal symmetry  was proposed  as a symmetry realized on superamplitudes in \cite{DHKS1} and proven for MHV and NMHV tree approximation superamplitudes.
Using recursion relations for superamplitudes~\cite{NimaParis,QMC1,nima}, the dual superconformal covariance of N$^k$MHV tree superamplitudes was then proven in \cite{QMC1}, and an explicit construction of ${\cal R}^{\rm MHV}_{n,0}$ in terms of dual conformal invariants was established in \cite{DH}. 
It is also interesting to note that conventional and dual superconformal symmetries combine into a Yangian structure \cite{DHP}.

It is well known that 1-loop $n$-point amplitudes in $\cn=4$ SYM can be expressed in terms of
scalar box integrals \cite{Bern:1994zx}.  The box integral expansion of 1-loop superamplitudes was initiated in \cite{DHKS1} and carried out systematically in \cite{DHKS2} using the method of maximal cuts \cite{BCF} to compute the amplitude. The dual conformal invariance of the ratio
${\cal R}^{{\rm NMHV}}_{n,1}$  was established for $n=6,7,8,9$ in \cite{DHKS2}.

\vspace{1mm}
\emph{Note Added:} 
After finishing this work, we were made aware (in recent email correspondence with Gabriele Travaglini) of similar work of Brandhuber, Heslop, and Travaglini \cite{BHT}.


\setcounter{equation}{0}
\section{N$^k$MHV tree superamplitudes and superconformal invariants}

We briefly review the needed tree level results.
We begin by giving the form of N$^k$MHV tree superamplitudes\footnote{We always omit the conventional energy-momentum $\d$-function $\d^{(4)}(\sum_i p_i^\m) $ although superamplitudes are covariant under dual symmetries only if the transformations of these $\d$-functions are included \cite{DHKS1}.}
 which are the `input data' for the 1-loop calculations:
\be \lab{kmhv0}
\ca_{n,0}^{{\rm N}^k{\rm MHV}}=   \frac {\d^{(8)}(\sum_{i=1}^n |i\> \h_{i}^a)}{ \cyc(1,\dots,n) }\cp^{{\rm N}^k{\rm MHV}}_{n,0}\,,
\qquad \cyc(1,\dots,n)\equiv\prod_{i=1}^n\<i,i+1\>\,.
\ee
Unless otherwise stated, all line labels are understood (mod $n$), i.e.~$i\equiv i+n$.
The $\ca_{n,0}^{{\rm N}^k{\rm MHV}}$ contain an  explicit $\d^{(8)}(\sum |i\>\h^a_i )$, which is a polynomial of degree 8 in the
$\h^{a}_i$,  while  $\cp^{{\rm N}^k{\rm MHV}}_{n,0}$ is a polynomial of degree $4k$.
 The complete
polynomial is of order $4(k+2).$

The MHV ($k=0$) superamplitude is particularly simple, with
$\cp^{ \rm MHV}_{n,0} =1$ \cite{nair}.
 Beyond the MHV level, several explicit representation of $\cp^{{\rm N}^k{\rm MHV}}_{n,0}$ for arbitrary $k$ and $n$ are known.\footnote{The MHV vertex expansion (first introduced in \cite{CSW}) provides an efficient representation of $\cp^{{\rm N}^k{\rm MHV}}_{n,0}$. The explicit MHV vertex representation for superamplitudes was presented, and its validity proven, in the recent papers~\cite{Elvang:2008na,Elvang:2008vz}, generalizing the earlier NMHV level analysis of~\cite{Georgiou:2004by,BEF}.
A novel version of the MHV vertex expansion based on supersymmetric shift recursion relations was derived in \cite{Kiermaier:2009yu}. In addition to these, there are also ambi-twistor representations of the tree amplitudes \cite{ArkaniHamed:2009si,Mason:2009sa,Hodges:2009hk}.}
The most convenient representation for the purposes of this paper,
is that of~\cite{DHKS1,DHKS2,DH} which expresses $\cp^{{\rm N}^k{\rm MHV}}_{n,0}$ in terms of dual superconformal invariants. At the NMHV level,
 this representation is 
\bea \lab{nmhv0}
\cp^{\rm NMHV}_{n,0}&=&\sum_{j=3}^{n-2}\sum_{k=j+2}^{n} R_{1jk}\, ,\\
R_{ijk} &\equiv&  \frac{\<j-1,j\>\<k-1,k\>\d^{(4)}(\Xi_{ijk})}{x_{jk}^2 \<i| x_{ij}x_{jk}|k\>\<i|x_{ij}x_{jk}|k-1\>  \<i| x_{ik}x_{kj}|j\>\<i|x_{ik}x_{kj}|j-1\> } \, ,\\[2.8mm]
\Xi_{ijk} &\equiv& \<i| x_{ij}x_{ij}|\theta_{ki} \> +  \<i| x_{ik}x_{kj}|\theta_{ji} \>\,.
\eea
The $R_{ijk}$ are the superconformal invariants of \cite{DHKS1} which depend on the differences
of zone variables $x_{ij} =x_i-x_j$, that are expressed as bispinors above, and on $|\theta_{ij}\> = |\theta_i\>-|\theta_j\>$. The $R_{ijk}$ and $\Xi_{ijk}$ are defined for labels $i,j,k$ satisfying $j-i\geq2$, $k-j\geq 2$, $k-i\geq1$. Spinor indices are suppressed unless needed for clarity.  At level
N$^k$MHV, the $\cp^{{\rm N}^k{\rm MHV}}_{n,0}$ are expressed in terms of more complicated superconformal invariants, see \cite{DH} and Sec.~2D of \cite{DSVW}. The arguments of \cite{DHKS1, QMC1, DH} show that the $\ca_{n,0}^{{\rm N}^k{\rm MHV}}$ transform covariantly under dual superconformal transformations.

One-loop superamplitudes constructed from unitary cuts contain products of tree subamplitudes which are either N$^k$MHV (with $k\geq0$) as discussed above or 3-point anti-MHV superamplitudes  of the form
\be \lab{anti}
  \ca_{3,0}^\text{anti-MHV}(i,j,k)  
  ~=~
  \frac{\delta^{(4)}\big( [ij] \eta_{ka} +[jk] \eta_{ia}
  + [ki] \eta_{ja}\big)}{[ij][jk][ki]} \,,
\ee
which is a polynomial of degree 4.  Effectively, a 3-point  anti-MHV subamplitude has $k=-1$, since it is a degree $4(k+2)=4$ Grassmann polynomial.


\setcounter{equation}{0}
\section{One-loop superamplitudes and box coefficients}\label{secboxes}

All 1-loop
amplitudes of $\cn = 4$ SYM can be expressed in terms of scalar box  integrals \cite{Bern:1994zx},
and the same is true of the superamplitudes in which they are packaged.  Our analysis therefore
starts with the representation
\be \lab{A1loop}
\ca_{n,1} = \sum_{r,s,t,u} c_{rstu}\, F_{rstu} \,.
\ee
Each term in the sum is a contribution of a box diagram
\begin{figure}
\begin{center}
 \includegraphics[height=5.5cm]{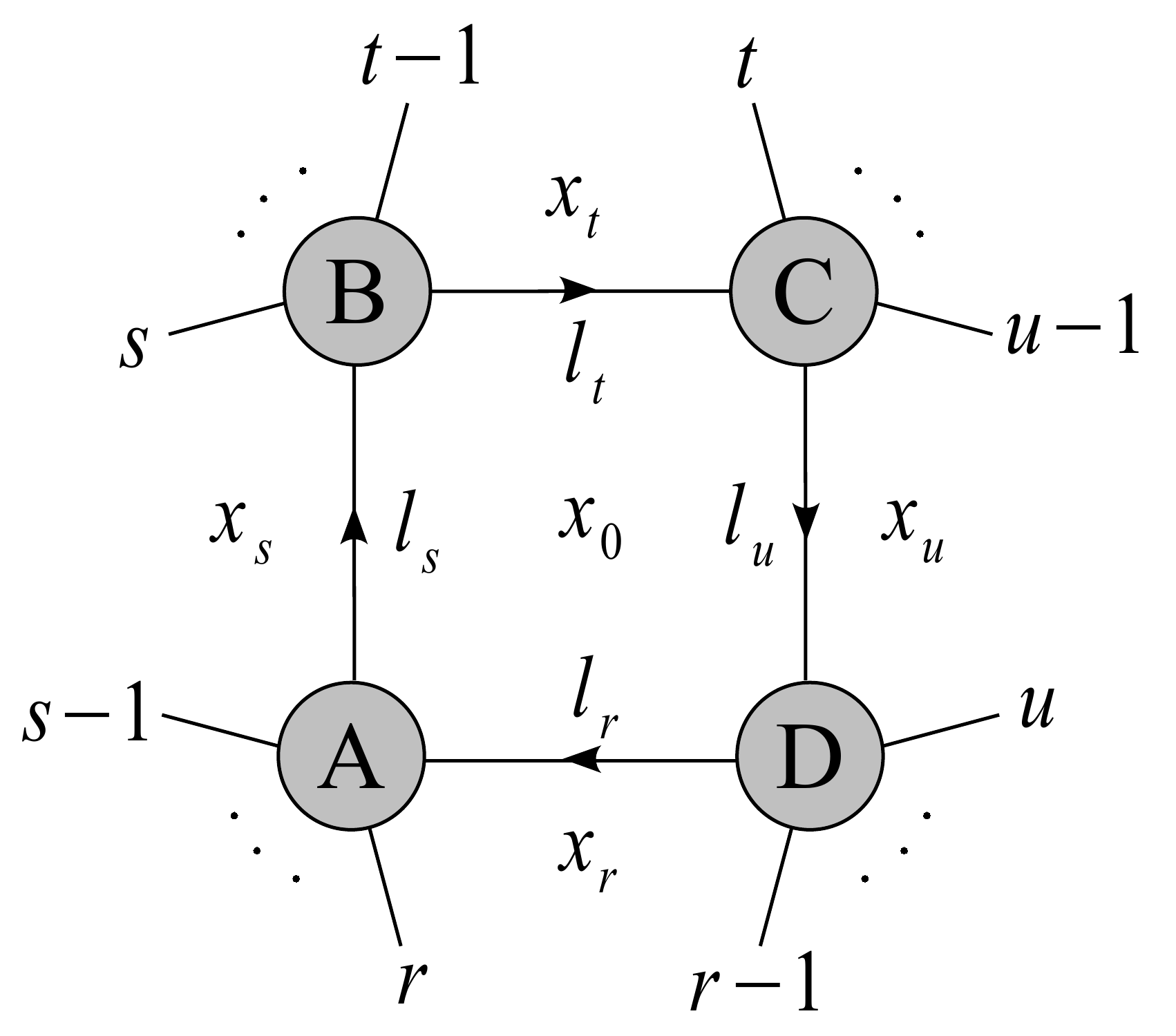}\\[3mm]
\end{center}
\vspace{0mm}
\caption{A general box diagram with tree subamplitudes $A$, $B$, $C$, $D$.  There is a zone variable $x_i$ between each pair of external lines. Four of these are indicated. The zone vector $x_0$ is discussed in Sec.~\ref{secDan}.}
\lab{fig:boxdiag}
\end{figure}
in which the first external line of subamplitude $A$, containing $n_A$ lines, is denoted by $r$, the first external line of subamplitude $B$, containing $n_B$ lines,  is denoted by $s$, etc.
 The conventions are indicated in Fig.~\ref{fig:boxdiag}.
 The full superamplitude  includes the contribution of all box diagrams with inequivalent partitions
$n_A+n_B+n_C+n_D=n+8$  and, for each of these, a sum over the $n$ cyclic permutations of the external lines.   We  define the `external masses' of the subamplitudes
by $K_A^2 =(p_r +p_{r+1}+\ldots p_{s-1})^2$, etc.  Box integrals are classified by the number of
non-vanishing  external masses,  $K_I^2 >0$ with $I=A,B,C,D$\,, which they contain. This classification will be given in detail in the next subsection.

For each box diagram one must integrate over the loop momentum $l$ and the $4\times 4$ $\h$ variables for the internal particles.  The maximal cut method \cite{BCF} gives an efficient way to separate these
tasks.  The $F_{rstu}$ are simply  1-loop Feynman integrals containing four scalar propagators, which are made dimensionless through the inclusion of an overall factor of $\D_{rstu}$, defined as\footnote{Note that $x_{st} =  K_B^2$ and $x_{su} = (K_B+K_C)^2$ etc.}
\be \lab{delta}
\D_{rstu} = \frac12 (x_{su}^2x^2_{rt}-x_{st}^2 x^2_{ur})\,.
\ee
 The $F_{rstu}$ are IR divergent when one or more subamplitudes are 3-point vertices. They
 are tabulated in Appendix B for various configurations of external lines. The box coefficients $c_{rstu}$ are obtained from the box diagrams with the four internal
lines cut,
\be \lab{boxcoff}
c_{rstu} =\frac{1}{ \D_{rstu} } \int\!\!\! \prod_{i=r,s,t,u}\!\!\!\! d^4\h^a_{l_i}\,\, \ca_{n_A} \ca_{n_B} \ca_{n_C}\ca_{n_D}\,.
\ee
  The $\h$-integrals produce a canceling
factor of the `Jacobian' $\D_{rstu}$,  and  the products $c\, F$ in \reef{A1loop} then do not contain any factors of $\D$.
The maximal cut conditions actually have two solutions $\cs_{\pm}$ for the momenta of the internal lines \cite{BCF}. For certain subamplitude configurations only  one of them  contributes. When both contribute we must modify \reef{boxcoff} by averaging over $\cs_{\pm}$.

 There is a simple and useful argument based on the order of the $\h$-polynomials to ascertain
 which subamplitudes contribute to  the 1-loop superamplitude  $\ca_{n,1}^{{\rm N}^k{\rm MHV}}$,
 which is an $\h$-polynomial of order $4(k+2)$.  The internal $\h$ integrals in \reef{boxcoff} remove $4\times 4 =16~ \h$'s. Thus,
 if the subamplitudes are N$^{k_I}$MHV,  we must have $  k_A + k_B + k_C + k_D = k-2$.

\subsection{Box coefficients $c_{rstu}$} \label{ss:boxes}

In this section we use the results of \cite{DHKS2} for  internal $\h$-integrals in
the box coefficients \reef{boxcoff}.  As first shown in \cite{BEF}, these integrals effectively automate the sum over the various internal particles of the $\cn=4$ theory which propagate in the loop.
For each distinct type of box integral, one must carefully analyze the product of contributing tree subamplitudes. There are two essential properties.
First, the Grassmann $\d^{(4)}$ and $\d^{(8)}$ factors of these subamplitudes are used to fix the values of the internal  $\h^a_{l_i}$. Secondly, at arbitrary N$^k$MHV level,\footnote{We use $\cp_I$ to denote the  $\cp_{n_I,0}$ factor of subamplitude $I$.} the
$\cp_I$ are chosen {\emph not}  to include the Grassmann variables associated with internal lines, i.e. $\eta_{l_r}, \eta_{l_s}, \eta_{l_t}$, and $\eta_{l_u}$. That this is possible can be seen from the explicit representation given in \cite{DSVW} and it is discussed in Sec.~\ref{secDan} below. As a consequence of these two properties, the $\eta_{l_i}$ integrals can be carried out trivially.

 It is common in all cases  that an overall factor of
$\ca_{n;0}^{\rm MHV}({\rm ext})$ naturally appears. Also, the various subamplitudes contain spinor products which combine into a factor of $\Delta_{rstu}$ and cancel the $\Delta_{rstu}^{-1}$ in the definition of $c_{rstu}$.
Let us summarize the results for $c_{rstu}$
for the several types of box diagrams,
with N$^{k_I}$MHV subamplitudes for general $k_I$:
\begin{itemize}
\item {\bf 4m:}
For the 4m box, each subamplitude must be MHV or higher, so this box  contributes only at the N$^2$MHV level and beyond.  We postpone the discussion until Sec.
\ref{secDan}. 

\item  {\bf 3m:}  There is now a  ``massless'' 3-point subamplitude which is placed at $A$.  It is
required by special kinematics to be anti-MHV. The box coefficient can be written as
\bea
  \label{the3m}
   c^{\rm 3m}_{rstu}
   = \ca_{n;0}^{\rm MHV} \; R_{rtu}  \,\cp_B\,\cp_C\,\cp_D \, .
\eea
One must keep the particle label (in)equalities  $s=r+1,  ~t >s+1,~ u >t+1,~r>u+1 $ in mind.
 All external line labels $i$ are defined mod $n$.

\item  {\bf 2mh:}   In this case there are massless 3-point subamplitudes at  $A$ and $B$.  Thus we have the (in)equalities  $s=r+1,~t=s+1=r+2,~ u>t+ 1,~r >u+1$.
There are two contributions which must be added: one with $A$ anti-MHV and $B$ MHV, the other with the opposite arrangement. It is clear that these two boxes are just extreme cases of the 3m boxes. We therefore have
\bea
  c^{\rm 2mh}_{rstu}
  = c^{\rm 3m}_{rstu} +  c^{\rm 3m}_{stur} = \ca_{n;0}^{\rm MHV}\,(R_{rtu} + R_{sur})\cp_C\cp_D\,.
\eea

\item  {\bf 2me:}
The massless subamplitudes $A$ and $C$ must be anti-MHV for generic external momenta. The result for the 2me box is
\bea
  \label{G2me}
   c^{\rm 2me}_{rstu}
   = \ca_{n;0}^{\rm MHV} \times \;
   \frac{1}{2} \sum_{\cs^\pm}
  \,\cp_B\,\cp_D \,
\eea
with $s=r+1,~ u=t+1, ~t>s+1$, and $r>u+1$.

\item  {\bf 1m:}
We take $A,~B,~C$ massless. There are two distinct contributions:
\begin{itemize}
\item [{\bf(a)}] $A$ and $C$ are anti-MHV. This is a special case of a 2me (with $\cp_B=1$).
\item[{\bf(b)}] $A$ and $C$ are MHV and $B$ is anti-MHV. This is a special case of a 3m box.
\end{itemize}
The result is then
\bea
  c^{\rm 1m}_{rstu}
  = c^{\rm 2me}_{rstu} +  c^{\rm 3m}_{stur}\,.
\eea
with $s=r+1$, $t=r+2$, $u=r+3$ and $r >u+1$.
\end{itemize}


\subsection{One-loop MHV amplitude}

It follows from $\eta$-counting that only the type (a) 1m boxes and  the 2me boxes contribute to the one-loop MHV amplitude.
The 1m box has $\cp_{B,D} =1$, so by \reef{G2me} we have
$c^{1m}_{rstu}=\ca_{n;0}^{\rm MHV}({\rm ext})$. Hence the 1m contributes
$\ca_{n;0}^{\rm MHV}\big(F^\M_{1234}$ + cyclic\big).

In the 2me box coefficient, $\cp_{B,D} = 1$, so $c^{\rm 2me}_{rstu}=c^{\rm 2me}_{r,r+1,t,t+1}=\ca_{n;0}^{\rm MHV}({\rm ext})$ with $t>r+2$ and $r> t+2$. The 2me contribution to the amplitude is therefore
$\ca_{n;0}^{\rm MHV}\big(\frac{1}{2}\sum_{t=4}^{n-2} F^\MMe_{12t,t+1}$ + cyclic\big). The 1/2 compensates the overcounting due to the symmetry of the 2me box diagram.

The full result for the 1-loop amplitude is then
\bea
  \label{MHV}
  \ca_{n;1}^\text{MHV}
 & =& \ca_{n;0}^{\rm MHV}\cp^{\rm MHV}_{n,1}\\
 \cp^{\rm MHV}_{n,1}&=&
   \Big( F^\M_{1234} + \frac{1}{2}\sum_{t=4}^{n-2} F^\MMe_{12t,t+1}
  + {\rm cyclic} \Big) \, .
\eea
Using the results for the box integrals $F_{rstu}$ in Appendix B one finds that the IR divergent parts $^\eps F_{rstu}$ satisfy a nice relationship, namely
\bea
  \label{IR2me3m}
  \frac{1}{2}\sum_{t=4}^{u} {^\eps} F^\MMe_{12t,t+1}
  = {^\eps} F^\MMM_{124,u+1}
 \qquad\text{ for }u\geq5
  \, .
\eea
Using \reef{IR2me3m}, the IR divergent part of the MHV amplitude can be written very compactly as
\bea
  {^\eps}\ca_{n;1}^\text{MHV}
  ~=~ \ca_{n;0}^{\rm MHV} \Big({^\eps}F^\M_{1234} + {^\eps}F^\MMM_{124,n-1}
  + {\rm cyclic}\Big)\,.
\eea
With the results for $F_{rstu}$ in appendix~\ref{appboxes}, and exploiting cyclic symmetry, one finds
\bea \lab{mhvir}
  {^\eps}\cp_{n;1}^\text{MHV}
  ~\equiv~
  {^\eps}\ca_{n;1}^\text{MHV}/
   \ca_{n;0}^{\rm MHV}
   ~=~
   - \frac{1}{\eps^2}\Big( (-x_{13}^2)^{-\eps} + {\rm cyclic}\Big)
  ~=~
  -
  \frac{1}{\eps^2} \sum_{i=1}^n (-s_{i,i+1})^{-\eps}
  \, ,
\eea
which is the `universal'  IR divergence of one-loop scattering amplitudes in $\cn =4$ SYM.


\setcounter{equation}{0}
\section {NMHV one-loop superamplitudes}

We have discussed the fact that one-loop amplitudes are IR divergent and that this spoils dual conformal symmetry.  On the other hand, the external state dependent IR singularity is encoded in a tree level factor:  it is captured fully by
 $\ca^\text{N$^k$MHV}_{n,0} \times{}^{\eps}\cp^{\rm MHV}_{n,1}$.  
For this reason, the dual conformal properties of the ratio
${\cal R}_{n,1}^{\rm NMHV}$ defined in \reef{ratio} were studied in \cite{DHKS2}.  Thus we expand the quantities in \reef{ratio} to first order in the 't Hooft coupling  $\l = g^2N/4\pi$  and define:
\bea
\ca^{\rm MHV}_n &=&  \ca^{\rm MHV}_{n,0} \big[ 1 +\l  \big( \cp^{\rm MHV}_{n,1} + O(\eps) \big)\big]\, ,\\[2mm]
\ca^{\rm NMHV}_n &=&  \ca^{\rm MHV}_{n,0}
\big[ 
\cp^{\rm NMHV}_{n,0}+ \l  \big(\cp^{\rm NMHV}_{n,1}+ O(\eps) \big)\big]\,.
\lab{P1}
\eea
The quantities $\cp_{n,1}$ are infrared divergent,  but the ratio
\be \lab{1lpratio}
{\cal R}^{\rm NMHV}_{n,1} =  \cp^{\rm NMHV}_{n,1} - \cp^{\rm NMHV}_{n,0} \cp^{\rm MHV}_{n,1}
\ee
must be IR finite. The claim is that it is also dual conformal invariant.  We will prove this for all $n$.

The subamplitude factors $\cp_I$ that can occur in the box coefficients of Sec.~\ref{ss:boxes} are restricted for one-loop NMHV amplitudes. This follows from the $\eta$-count given below \reef{boxcoff}.
For the  3m box, subamplitudes $B$, $C$ and $D$ are all MHV, so $\cp_{B,C,D}=1$, and hence
 $c_{rstu}^{3m} = R_{rtu}$.  
The 2me box requires considerably more intricate arguments, since one of
the massive subamplitudes is NMHV. As shown very nicely in \cite{DHKS2}, the 2me coefficient can be written as a certain sum over 3m coefficients.
 We summarize the results for all needed box coefficients:
\begin{itemize}
\item
  $c^{\rm 3m}_{rstu} = \ca_{n;0}^{\rm MHV}\,R_{rtu}$ with $s=r+1$.
\item
  $c^{\rm 2mh}_{rstu} =c^{\rm 3m}_{rstu} +c^{\rm 3m}_{stur}
    = \ca_{n;0}^{\rm MHV}(R_{rtu} + R_{sur})$ with $s=r+1$ and $t=r+2$.
\item
  $c^{\rm 2me}_{rstu} =
   \sum_{v=u}^{r-3}\sum_{w=v+2}^{r-1}
  c^{\rm 3m}_{r,r+1,vw}  =  \ca_{n;0}^{\rm MHV}
  \sum_{v=u}^{r-3}\sum_{w=v+2}^{r-1}
  R_{rvw}$.\\[2mm]
 This takes $D$ to be NMHV and $B$ to be MHV. The reverse arrangement is related by cyclic symmetry and is thus accounted for after we sum over all cyclic permutations. Note that the sum over $v,w$ includes only the \emph{external} legs on the subamplitude $D$.
\item
  $c^{\rm 1m}_{rstu}
  = c^{\rm 2me}_{rstu} + c^{\rm 3m}_{stur}
  =
  \sum_{v=u}^{r-3}\sum_{w=v+2}^{r-1}
  c^{\rm 3m}_{r,r+1,vw}
  + c^{\rm 3m}_{stur}$,\\[2mm] \text{\ie}~
  $c^{\rm 1m}_{rstu}
  =  \ca_{n;0}^{\rm MHV}(
  \sum_{v=u}^{r-3}\sum_{w=v+2}^{r-1}
  R_{rvw} + R_{sur})$~
with $s=r+1$, $t=r+2$ and $u=r+3$.
\end{itemize}

Every box coefficient is a linear combination of the $R_{ijk}$-invariants. The one-loop superamplitude is a sum over all contributing boxes, $c\, F$. We find that $\cp^{\rm NMHV}_{n;1}$ defined in
\reef{P1} is
{
\bea
  \nonumber
  \cp^{\rm NMHV}_{n;1}
    &=&
  \sum_{s=4}^{n-3} \sum_{t=s+2}^{n-1}
   R_{1st}\, F^\MMM_{12st}
   +
   \sum_{t=5}^{n-1} R_{13t}\, F^\MMh_{123t}
   +
   \sum_{s=4}^{n-2} R_{1sn}\, F^\MMh_{n12s}
   +
   \sum_{k=4}^{n-3}
   \biggl[\,\sum_{s=k+1}^{n-2}\sum_{t=s+2}^n
   R_{1st} \biggr]\,  F^\MMe_{12k,k+1} \\[2mm]
   &&
   ~~~
   + R_{13n} \, F^\M_{n123}
   + \biggl[\,\sum_{s=4}^{n-2}\sum_{t=s+2}^n R_{1st} \biggr]\,  F^\M_{1234}
   ~~+~ {\rm cyclic} \, .
   \label{Pn1}
\eea}
It is useful to reorder the  sums so that  each individual $R_{1st}$ is multiplied by linear combinations of  box integrals, to wit
\begin{equation}\label{Pn1Rs}
\begin{split}
  \cp^{\rm NMHV}_{n;1}
  =
  &
  \sum_{t=5}^{n-1}  F^\MMh_{123t}\,R_{13t}+  F^\M_{n123}\,R_{13n}+
  \sum_{s=4}^{n-3} \sum_{t=s+2}^{n-1}
   \bigg[
   F^\M_{1234} + F^\MMM_{12st} + \sum_{k=4}^{s-1} F^\MMe_{12k,k+1}
   \bigg]\,R_{1st}\\[2mm]
   &
   +
   \sum_{s=4}^{n-2} \bigg[ F^\M_{1234}+ F^\MMh_{n12s}
   + \sum_{k=4}^{s-1} F^\MMe_{12k,k+1} \biggr]\,R_{1sn}
   ~~+~ {\rm cyclic}
  \, .
\end{split}
\end{equation}
When $s=4$ in the first line, the sum over $k$ is empty and we set it to zero.

Finally we need  to subtract the universal IR divergence in order to obtain the ratio  ${\cal R}^{\rm NMHV}_{n,1}$ as defined in~(\ref{1lpratio}). The result is a compact representation of the ratio:
\begin{equation}\label{remNMHV}
    \cR_{n,1}^\text{NMHV}=\sum_{s=3}^{n-2}\sum_{t=s+2}^n{\cal T}_{1st} R_{1st} ~+~\text{cyclic}\,,
\end{equation}
with
\begin{equation}
\begin{split}
  {\cal T}_{13t}&=
  F^\MMh_{123t}-F^\M_{1234}-\frac{1}{2}\sum_{u=4}^{n-2} F^\MMe_{12u,u+1}\hskip3cm\text{ for }~~5\leq t\leq n-1\,,
  \\[1mm]
  {\cal T}_{13n}&=F^\M_{n123}-F^\M_{1234}-\frac{1}{2}\sum_{u=4}^{n-2} F^\MMe_{12u,u+1}\,,\\[1mm]
  {\cal T}_{1st}&=
   F^\MMM_{12st}\, + \sum_{k=4}^{s-1} F^\MMe_{12k,k+1}-\frac{1}{2}\sum_{u=4}^{n-2} F^\MMe_{12u,u+1}
   \hskip1.8cm\text{ for }~~4\leq s,t\leq n-1\,,\\[1mm]
  {\cal T}_{1sn}&=F^\MMh_{n12s}\,+ \sum_{k=4}^{s-1} F^\MMe_{12k,k+1}-\frac{1}{2}\sum_{u=4}^{n-2} F^\MMe_{12u,u+1}
   \hskip1.7cm\text{ for }~~4\leq s\leq n-2\,.
\end{split}
\end{equation}

Unfortunately, none of the coefficients ${\cal T}_{1st}$ are dual conformal invariant,
and all except ${\cal T}_{14,n-1}$ are also infrared divergent.\footnote{IR finiteness of ${\cal T}_{14,n-1}$ follows from \reef{IR2me3m}.}
Thus it is not clear from the representation \reef{remNMHV} that $\cR_{n,1}^\text{NMHV}$ is dual conformal invariant --- or even IR finite!  We now proceed to bring \reef{remNMHV} to a manifestly dual conformal invariant form. The IR finite coefficient ${\cal T}_{14,n-1}$ will play a special role in our analysis.


\setcounter{equation}{0}
\section{Proof of dual conformal invariance}

The main tools for our manipulations of $\cR_{n,1}^\text{NMHV}$ are cyclic symmetry and the fact that there exist linear relations among the $R_{ijk}$-invariants. Our strategy, which is similar to that in \cite{DHKS2},  is to use these  properties to eliminate certain $R_{rst}$ from the sum
in \reef{remNMHV} and thus obtain a new representation of  $\cR^\text{NMHV}_{n;1}$
in which the combinations of box integrals that appear are both IR finite and dual conformal invariant.  Two steps are required to find this representation. First, we use the cyclic sum in~(\ref{remNMHV}) to make one term in the sum manifestly cyclically invariant.  Then we
systematically use linear relations between the $R_{ijk}$  to reduce the sum  to a smaller set of contributing  $R$-invariants.

\subsection{Using cyclicity}
The overall cyclic symmetry of~(\ref{remNMHV}) allows us
to eliminate $R_{14,n-1}$
 in favor of the cyclically invariant $R_{\rm tot}$, defined as
\begin{equation}\label{Rtot}
    R_{\rm tot}\equiv\cp^\text{NMHV}_{n;0}=\sum_{s=3}^{n-2}\sum_{t=s+2}^n R_{1st}\,.
\end{equation}
We choose to eliminate $R_{14,n-1}$ because its coefficient ${\cal T}_{14,n-1}$ is already IR finite. This step leads to the modified representation
\begin{equation}\label{remNMHV2}
    \cR_{n,1}^\text{NMHV}={\cal S}_{\rm tot}R_{\rm tot}+\sum_{s=3}^{n-2}\sum_{t=s+2}^n({\cal T}_{1st}-{\cal T}_{14,n-1}) R_{1st}~+~\text{cyclic}\,,\qquad
    {\cal S}_{\rm tot}=\frac{1}{n}\,\Bigl({\cal T}_{14,n-1}\,+\,\text{cyclic}\Bigr)\,.~~
\end{equation}
Note that $R_{14,n-1}$ now no longer appears in $\cR_{n,1}^\text{NMHV}$
because its coefficient vanishes.\footnote{
For the special case $n=6$, there is no $R_{14,n-1}$ in~(\ref{remNMHV}), and
(\ref{remNMHV2}) is then
to be understood with ${\cal T}_{14,n-1}\to 0$.}
Although it is far from obvious,  $\cs_{\rm tot}$ is our first dual conformal invariant combination of box integrals.
It will be expressed in terms of dual cross-ratio variables in
\reef{Stotuu} below.


\subsection{Eliminating `dependent' $R_{rst}$}

We now eliminate all $R_{13t}$ with $5\le t \le n$ and all $R_{14t}$ with
$ 6\le t \le n-2$ plus their  cyclically related companions.      We have already replaced $R_{14,n-1}$ and its cyclic
companions by $R_{\rm tot}$, which eliminates $n-1$ additional $R$-invariants.
  So in total,  we
eliminate  $2n^2-10n-1$ distinct $R_{ijk}$. We first eliminate $R_{13t}$, then in a separate step $R_{14t}$.

To eliminate the $R_{13t}$, we use the identity \cite{DHKS1,DHKS2}
 \begin{equation}\label{linrels}
     R_{i,i+2,j} = R_{i+2,j,i+1}\,,
 \end{equation}
 which implies
\begin{equation}\label{13t}
\begin{split}
    R_{13t} &=R_{3t2} = \PP^{2} R_{1,t-2,n}\qquad \qquad \qquad\text{ for}\quad 6\leq t\leq n\,,\\
    R_{135} &=R_{352} = R_{524} = \PP^{4} R_{1,n-2,n}\,.
\end{split}
\end{equation}
The operator $\PP^k$  is defined as the shift $i \to i + k$ of all indices of the object immediately
to its right.  Applying~(\ref{13t}) to the ratio~(\ref{remNMHV2}), we obtain
\begin{equation}\label{remNMHV3}
\begin{split}
    \cR_{n,1}^\text{NMHV}\!=&\,{\cal S}_{\rm tot}R_{\rm tot}\!+\!
    \sum_{s=4}^{n-3}\sum_{t=s+2}^{n-1}
    \Bigl({\cal T}_{1st}-{\cal T}_{14,n-1}\Bigr) R_{1st}
    +\!\sum_{s=4}^{n-3}\Bigl({\cal T}_{1sn}+\PP^{-2}{\cal T}_{13,s+2}\!-\!(1+\PP^{-2}){\cal T}_{14,n-1}\Bigr) R_{1sn}\\[.5ex]
    &+\Bigl({\cal T}_{1,n-2,n}+\PP^{-2}{\cal T}_{13n}+\PP^{-4}{\cal T}_{135}-(1+\PP^{-2}+\PP^{-4}){\cal T}_{14,n-1}\Bigr) R_{1,n-2,n}
    ~~+~\text{cyclic}\,.
\end{split}
\end{equation}
Here, we have used the overall cyclic sum to convert shift operators $\PP^{q}$  acting on $R_{1st}$ invariants into shift operators $\PP^{-q}$ acting on scalar box integrals in the coefficients ${\cal T}_{1st}$\,.

The steps taken so far are essentially all that is needed to establish dual conformal symmetry for the cases $n=6,7$.
In Appendix~\ref{app67}  we show that the representation \reef{remNMHV3} agrees with the forms given in \cite{DHKS2}.
For the rest of this section we assume that $n \ge 8$.

For $n\geq8$, the coefficients in the representation~(\ref{remNMHV3}) of the ratio function are {\em not} all finite and dual conformal invariant. For example, all invariants $R_{14k}$ with $6\leq k\leq n-2$ have coefficients with IR singularities:
\begin{equation}
    {\cal T}_{14k}-{\cal T}_{14,n-1}
\,=\, F^\MMM_{124k}-F^\MMM_{124,n-1}
  \,=\,
  -\frac{1}{2\epsilon^2}
  \Bigl[(-x_{2k}^2)^{-\epsilon}+(-x_{1,n-1}^2)^{-\epsilon}-(-x_{1k}^2)^{-\epsilon}-(-x_{2,n-1}^2)^{-\epsilon}\Bigr]+\text{finite}\,.
\end{equation}
Fortunately, we can also eliminate all these $R_{14k}$ with IR-divergent coefficients  from the ratio~(\ref{remNMHV3}). Indeed, the
identities\footnote{Considerable evidence for this identity was given in \cite{DHKS2}. It was proven in the very recent \cite{BHT}.}
\begin{equation}\label{sumid}
  \sum_{s=3}^{k-2}\sum_{t=s+2}^k R_{1st}
  =  \sum_{s=2}^{k-3}\sum_{t=s+2}^{k-1}  R_{kst}
\end{equation}
can be used in conjunction with~(\ref{13t}) to eliminate all $R_{14k}$ ($6\leq k\leq n-2$)  without reintroducing the already eliminated invariants $R_{13t}$ or $R_{14,n-1}$. To see this, we proceed as follows.
For $k=6$, we solve~(\ref{sumid}) for $R_{146}$ and obtain
\begin{equation}\label{k=6}
    R_{146}=R_{624}+R_{625}+R_{635}-R_{135}-R_{136}=\PP^5\bigl(R_{1,n-3,n-1}+R_{1,n-3,n}+R_{1,n-2,n}\bigr)-\PP^4 R_{1,n-2,n}-\PP^2 R_{14n}\,.
\end{equation}
For $k=7$, we solve~(\ref{sumid}) for $R_{147}$ and use~(\ref{k=6}) to eliminate $R_{146}$. This gives
\begin{equation}\label{k=7}
\begin{split}
    R_{147} &=\PP^6\bigl(R_{1,n-4,n-2}+R_{1,n-4,n-1}+R_{1,n-4,n}+R_{1,n-3,n-1}+R_{1,n-3,n}+R_{1,n-2,n}\bigr)\\
    &\quad-\PP^5\bigl(R_{1,n-3,n-1}+R_{1,n-3,n}+R_{1,n-2,n}\bigr)-\PP^2 R_{15n}-R_{157}\,.
\end{split}
\end{equation}
This approach can be generalized to all $k\geq6$. We solve~(\ref{sumid}) for $R_{14k}$, and eliminate all $R_{14t}$ with $t<k$ using the previously obtained identities.
An inductive argument then shows that, for the full range $6\leq k\leq n-2$,
\begin{equation}\label{14k}
    R_{14k}=\PP^{k-1}\Biggl[\,\sum_{s=n-k+3}^{n-2}\,\sum_{t=s+2}^{n} R_{1st}\Biggr]-
    \PP^{k-2}\Biggl[\,\sum_{s=n-k+4}^{n-2}\,\sum_{t=s+2}^{n} R_{1st}\Biggr]-\PP^2R_{1,k-2,n}-\sum_{s=5}^{k-2}R_{1sk}\,.~
\end{equation}
We note that none of the (cyclic permutations of) $R_{13t}$ or
$R_{14t}$ that we want to eliminate appear on the right hand side
of~(\ref{14k}).

\subsection{Manifestly dual conformal invariant amplitude}

Using~(\ref{k=6}) and~(\ref{14k}) to eliminate all $R_{14k}$ with $6\leq k\leq n-2$ from $\cR_{n,1}^\text{NMHV}$ in~(\ref{remNMHV3}), we obtain our final form for the ratio:
\begin{equation}\label{RwithS}
    \cR_{n,1}^\text{NMHV}={\cal S}_{\rm tot}R_{\rm tot}+{\cal S}_{14n}R_{14n}+\sum_{s=5}^{n-2} \sum_{t=s+2}^{n} {\cal S}_{1st} R_{1st}~~+~\text{cyclic}\,,
\end{equation}
with
\begin{equation}\label{SofF}
\begin{split}
      {\cal S}_{\rm tot}&~=~ \frac{1}{n}\biggl(F^\MMM_{124,n-1}-\frac{1}{2}\sum_{u=4}^{n-2} F^\MMe_{12u,u+1}~+~\text{cyclic}\biggr)\,,\\
    {\cal S}_{1st}
    &~=~
    \sum_{u=4}^{s-1} F^\MMe_{12u,u+1}\,+F^\MMM_{12st}-F^\MMM_{124t}+\sum_{k=n+3-s}^{n-2}\!\!\PP^{1-k}\bigl(F^\MMM_{124k}-F^\MMM_{124,k+1}\bigr)
   \hskip.7cm\text{ for }~~5\leq s,t\leq n-1\,,\\
    {\cal S}_{1sn}&~=~\sum_{u=4}^{s-1} F^\MMe_{12u,u+1}\,+
    F^\MMh_{n12s}-F^\MMM_{124,n-1}
    +\PP^{-2}\,  \bigl(-F^\M_{1234}+F^\MMh_{123,s+2} -F^\MMM_{124,s+2} \bigr)\,
    \\
   &\qquad+\sum_{k=n+3-s}^{n-2}\!\!\PP^{1-k}
    \bigl(F^\MMM_{124k}-F^\MMM_{124,k+1}\bigr)
   \hskip5.4cm\text{ for }~~4\leq s\leq n-3\,,\\
      {\cal S}_{1,n-2,n} &~=~
       \sum_{u=4}^{n-3} F^\MMe_{12u,u+1}\,+F^\MMh_{n12,n-2}-F^\MMM_{124,n-1}+ \PP^{-2}  \bigl( F^\M_{n123} - F^\M_{1234}-F^\MMM_{124,n-1}\bigr)\\
       &\qquad+ \PP^{-4}  \bigl(-F^\M_{1234}+F^\MMh_{1235}-F^\MMM_{1246} \bigr)+\sum_{k=6}^{n-2}\PP^{1-k}\bigl(F^\MMM_{124k}-F^\MMM_{124,k+1}\bigr)\,.
\end{split}
\end{equation}
In all coefficients ${\cal S}$, empty sums are understood to vanish.

All ${\cal S}$ coefficients in~(\ref{SofF}) are finite and dual conformal invariant.  To verify this
we checked
  that the infrared divergences cancel in
 all coefficients, and we then showed that each of them is invariant
under the conformal inversion, which acts on zone variables or invariant squares of their differences  as
\be \lab{invx}
I[x_{\a\dot{\b}}] = x_{\b\dot{\a}}/x^2\,, \qquad\qquad I[x_{ij}^2]= x_{ij}^2/(x_i^2 x_j^2)\,.
\ee
Inversion symmetry guarantees that all  ${\cal S}$ coefficients can be expressed as functions of
the dual conformal invariant cross-ratios:
\be \lab{xrat}
u_{ijkl} =  \frac{x_{ik}^2x_{jl}^2}{x_{il}^2 x_{jk}^2}\,.
\ee
In the next section we list these expressions and indicate briefly how they were obtained. In particular, we use the di-log identity
\bea
  \Li_2\!\big(1 - z\big) + \Li_2\!\Big(1 - \frac{1}{z}\Big)
   = -\frac{1}{2} (\log z)^2 \, .
\eea

\subsection{Coefficients in terms of dual conformal cross-ratios}

To express the coefficients ${\cal S}_{1st}$ with $5\leq s,t\leq n-1$ in terms of dual conformal cross-ratios, we first compute
\bea
   \nonumber
    &&\hspace{-5mm }{\cal S}_{15t}
    ~=~
   \Li_2\!\left(\!1\!-u_{12t4}\right)
    -\,\Li_2\left(\!1\!-u_{1254}\right)
   \!-\!\Li_2\!\left(\!1\!-
   u_{12t5}
   \right)
   \!-\Li_2\!\left(1\!-\!
   u_{1745}
   \right)
   \!+\Li_2\!\left(1\!-\!
   u_{2745}
   \right)  \\[.7ex]
   \nonumber
   &&\quad\quad\quad
   -\tfrac{1}{2}\ln\left(
   u_{1275}
   \right)\ln\left(
   u_{1745}
   \right)
   -\tfrac{1}{2}\ln\left(
   u_{12t5}
   \right)\ln\left(
   u_{1t45}
   \right)
   -\tfrac{1}{2}\ln\left(
   u_{24t7}
   \right)\ln\left(
   u_{1254}
   \right)\, , \\[2.4ex]
      \nonumber
      &&\hspace{-5mm }
    {\cal S}_{1,s+1,t}-{\cal S}_{1st}
    ~=~F^\MMe_{12s,s+1}+F^\MMM_{12,s+1,t}-F^\MMM_{12st}+F^\MMM_{s,s+1,s+3,1}-F^\MMM_{s,s+1,s+3,2}\\[1.5ex]
   \nonumber
    &&=~
   \Li_2\!\left(\!1\!-u_{12ts}\right)
    -\Li_2\left(\!1\!-u_{21s,s+1}\right)
   \!-\!\Li_2\!\left(\!1\!-
   u_{12t,s+1}
   \right)
   \!-\!\Li_2\!\left(1\!-\!
   u_{1,s+3,s,s+1}
   \right)\\[.7ex]
   &&\quad
      \nonumber
   ~+\!\, \Li_2\!\left(1\!-\!
   u_{2,s+3, s,s+1}
   \right)
   -\tfrac{1}{2}\ln\left(
   u_{12,s+3,s+1}
   \right)\ln\left(
   u_{1,s+3,s,s+1}
   \right)
   -\tfrac{1}{2}\ln\left(
   u_{12t,s+1}
   \right)\ln\left(
   u_{1ts,s+1}
   \right)\\[.7ex]
   &&\quad
   ~ -\tfrac{1}{2}\ln\left(
   u_{2st,s+3}
   \right)\ln\left(
   u_{21s,s+1}
   \right) \, ,
  \qquad \qquad \qquad\text{ for }5\leq s,t\leq n-1\,,
\eea
which can then be summed to obtain
\bea
   \nonumber
    {\cal S}_{1st}
    &=&\Li_2\!\left(\!1\!-
    u_{12t4}
    \right)
    \!-\!\Li_2\!\left(\!1\!-
    u_{12ts}
    \right)
    +\sum_{i=5}^s\Bigl[
   \Li_2\!\left(1\!-\!
   u_{2,i+2,i-1,i}
   \right)
    -\Li_2\left(\!1\!-
    u_{12i,i-1}
    \right)
   \\[0.2ex]&&   \nonumber
   ~
   -\, \Li_2\!\left(1\!-\!
   u_{1,i+2,i-1,i}
   \right)
   -\tfrac{1}{2}\ln\left(
   u_{21i,i+2}
   \right)\ln\left(
   u_{1,i+2,i-1,i}
   \right)
   -\tfrac{1}{2}\ln\left(
   u_{12ti}
   \right)\ln\left(
   u_{1t,i-1,i}
   \right)\\[.7ex]
   &&~-\tfrac{1}{2}\ln\left(
   u_{2,i-1,t,i+2}
   \right)\ln\left(
   u_{12i,i-1}
   \right)\Bigr] \, ,
   \qquad \qquad \qquad\text{ for }5\leq s,t\leq n-1\,.
\eea

A short computation shows that ${\cal S}_{14n}$ can be expressed in terms of dual conformal cross-ratios as
\begin{equation}
\begin{split}
    {\cal S}_{14n}
    &= \Li_2\left(1-
    u_{24n,n-1}
    \right)
   +\,\Li_2\left(1-
   u_{214,n-1}
   \right)
  +\ln\left(
  u_{24n,n-1}
  \right)
  \,\ln\left(
  u_{214,n-1}
  \right)-\tfrac{\pi^2}{6}\,.
\end{split}
\end{equation}
Using
\begin{equation}
\begin{split}
  {\cal S}_{1sn}-{\cal S}_{1s,n-1}&=-F^\M_{n-1,n12}+F^\MMh_{n-1,n1s}+F^\MMh_{n12s}-F^\MMM_{12s,n-1}-F^\MMM_{n-1,n2s}\\[1mm]
  &=\Li_2\left(1-
  u_{21s,n-1}
  \right)
  +\Li_2\left(1-
  u_{2sn,n-1}
  \right)
  +\ln\left(u_{2sn,n-1}\right)\,\ln\left(u_{21s,n-1}\right)
  -\tfrac{\pi^2}{6}
\end{split}
\end{equation}
for $5\leq s\leq n-3$,
we then have
\bea\label{S1sn}
  \nonumber
  \hspace{-0.3mm}{\cal S}_{1sn}
  \hspace{-3mm}&=&\hspace{-3mm}
   \Li_2\left(1-  u_{2sn,n-1}   \right)
    +\Li_2\!\left(\!1\!-u_{214,n-1} \right)
    +\ln\left(u_{2sn,n-1}\right)\,\ln\left(u_{21s,n-1}\right) \\[1mm]
    \nonumber
    &&
    \hspace{-4mm}+\sum_{i=5}^s\Bigl[
   \Li_2\!\left(1\!-\! u_{2,i+2,i-1,i} \right)
    -\Li_2\left(\!1\!-u_{12i,i-1}
    \right)
   \!-\!\Li_2\!\left(1\!-\! u_{1,i+2,i-1,i}
   \right)
   -\tfrac{1}{2}\ln\left( u_{21i,i+2}
   \right)\ln\left( u_{1,i+2,i-1,i}
   \right)\\
   &&\hspace{-4mm}
   -\tfrac{1}{2}\ln\left( u_{12,n-1,i}
   \right)\ln\left(u_{1,n-1,i-1,i}\right)
   -\tfrac{1}{2}\ln\left(u_{2,i-1,n-1,i+2}\right)
   \ln\left(u_{12i,i-1}\right)\Bigr]-\tfrac{\pi^2}{6}
\eea
for the entire range $4\leq s\leq n-3$.

To obtain ${\cal S}_{1,n-2,n}$, we first compute
\bea
  \nonumber
  &&\hspace{-6mm}{\cal S}_{1,n-2,n}-{\cal S}_{1,n-3,n}\\[1.3ex]
    \nonumber
  &&= F^\MMe_{12,n-3,n-2}+F^\MMh_{n12,n-2}-F^\MMh_{n12,n-3}+ \PP^{-2}  ( F^\M_{n123}-F^\MMh_{123,n-1}) + \PP^{-4}  ( -F^\M_{1234}+F^\MMh_{1235} -F^\MMM_{1246})\\[.5ex]
  &&=
  \Li_2\left(1-u_{2n,n-3,n-2} \right)
  -\Li_2\left(1-u_{12,n-2,n-3}\right)
- \ln\left(u_{n1, n - 3, n - 2}\right)\ln\left(\frac{u_{1 2, n - 2, n - 1}}{u_{2, n - 3, n - 1, n}}\right) \,. \hspace{-13mm}
\eea
Combining this with ${\cal S}_{1,n-3,n}$ from~(\ref{S1sn}) gives
\bea
  \nonumber
  \hspace{-0.1mm}{\cal S}_{1,n-2,n}
  &=& \hspace{-0.2cm}\Li_2\left(1- u_{2n,n-3,n-2}  \right)-   \Li_2\left(1- u_{12,n-2,n-3} \right)
    +\Li_2\left(1-  u_{2,n-3,n,n-1}   \right)
    \!+\!\Li_2\!\left(\!1\!-u_{214,n-1} \right)    \\[1mm]
      \nonumber
   &&\hspace{-1.5cm}
    - \ln\left(u_{n1, n - 3, n - 2}\right)\ln\left(\frac{u_{1 2, n - 2, n - 1}}{u_{2, n - 3, n - 1, n}}\right)
    +\ln\left(u_{2,n-3,n,n-1}\right)\,\ln\left(u_{21,n-3,n-1}\right)
    \\[-.3ex]
  \nonumber
   &&\hspace{-1.5cm}
    +\sum_{i=5}^{n-3}\Bigl[
   \Li_2\!\left(1\!-\! u_{2,i+2,i-1,i}\right)
    -\,\Li_2\left(\!1\!-u_{12i,i-1}
    \right)
   \!-\!\Li_2\!\left(1\!-\! u_{1,i+2,i-1,i}
   \right)
   -\tfrac{1}{2}\ln\left( u_{21i,i+2}
   \right)\ln\left( u_{1,i+2,i-1,i}
   \right)\\[.5ex]
   \nonumber
  &&\hspace{-1.5cm}
   -\tfrac{1}{2}\ln\left( u_{12,n-1,i}
   \right)\ln\left(u_{1,n-1,i-1,i}
   \right)
   -\tfrac{1}{2}\ln\left(u_{2,i-1,n-1,i+2}
   \right)\ln\left(u_{12i,i-1}
   \right)\Bigr] -\tfrac{\pi^2}{6} \,.
\eea

Finally, we turn to ${\cal S}_{\rm tot}$. Due to the cyclic sum in its definition~(\ref{remNMHV2}), its expression in terms of cross-ratios is more complicated than for all other ${\cal S}$. The cases $n=7,8$ are special. The ratio for $n=7$ is discussed in appendix~\ref{app67}, and ${\cal S}^{(n=7)}_{\rm tot}$ is presented there.  For $n=8$, we find
\begin{equation}
8\, {\cal S}^{(n=8)}_{\rm tot}
  =\,
  -\,\Li_2\big(1-  u_{1247}^{-1}
  \big)
  +\frac{1}{2}\sum_{u=4}^{6}\Li_2\big(1-u_{12u,u+1}^{-1}
  \big)
  +\frac{1}{4}\ln\left(u_{8145}\right)\ln\left(\frac{u_{1256}u_{3478}}{u_{2367}}\right)
 ~+~ {\rm cyclic}\, .~~
\end{equation}
For general $n$, we are lead to the following expression
\begin{equation}\lab{Stotuu}
\begin{split}
  n{\cal S}_{\rm tot}
  &=
  -\Li_2\big(1- u_{124,n-1}^{-1}
  \big)
  +\frac{1}{2}\sum_{u=4}^{n-2}\Li_2\big(1-u_{12u,u+1}^{-1}
  \big)
  \\[.2ex]
  &
  +\tfrac{1}{2} \ln\left(u_{n134}\right) \, \ln\left(u_{136,n-2}\right)
  -\tfrac{1}{2}\ln\left(u_{n145}\right) \ln\left(u_{236,n-2}\,u_{2367}\right)
  ~+~ {\mathbf{s}_n} ~+~ {\rm cyclic}\,,
\end{split}
\end{equation}
where ${\mathbf{s}_n}$ is defined separately for odd/even $n$:
\bea
  n~\text{odd:}~~~
  \mathbf{s}_n\hspace{-2mm}
  &=&\hspace{-2mm}
  \sum_{i=4}^{\frac{n-1}{2}} \ln\left(u_{n1i,i+1}\right)
   \sum_{j=1}^{i-1} \ln\left(u_{j,j+1,i+j,n-i+j}\right) \, , \\[.2ex]
  n~\text{even:}~~~
  \mathbf{s}_n
  \hspace{-2mm}&=&\hspace{-2mm}
  \sum_{i=4}^{\frac{n-2}{2}} \ln\left(u_{n1i,i+1}\right)
   \sum_{j=1}^{i-1} \ln\left(u_{j,j+1,i+j,n-i+j}\right)
   \!+\!\frac{1}{4} \ln\left(u_{n1,\frac{n}{2},\frac{n}{2}+1}\right)
   \sum_{i=1}^{\frac{n-2}{2}}
   \ln\left(u_{i,i+1,i+\frac{n}{2},i+\frac{n}{2}+1}\right)
   \, . ~\nonumber
\eea
To prove these expressions, one exploits cyclic symmetry to collapse the double sum in
$\mathbf{s}_n$ to a single sum plus boundary terms. We spare the reader the details of
this proof.

This completes our derivation of a manifestly IR finite and dual conformal invariant expression for the
ratio $\cR_{n,1}^\text{NMHV}$ of the one-loop $n$-point NMHV amplitude.

\setcounter{equation}{0}
\section{Superconformal covariance of box coefficients for \newline all one-loop N$^k$MHV superamplitudes}\label{secDan}

Individual box integrals $F_{rstu}$ are IR divergent and their finite parts are not manifestly dual conformal invariant.
Therefore it required an intricate argument to obtain a representation of the
NMHV ratio $\cR_{n,1}^{\rm NMHV}$ which contains \emph{linear combinations} of the $F_{rstu}$ in which
undesired singular and non-invariant terms cancel.   The `box coefficients'
\reef{boxcoff}, however,  which are obtained directly from the maximal cut of the superamplitude,  have much better properties.  The coefficient of each individual box diagram is covariant
under superconformal transformations. This was proven for NMHV superamplitudes in \cite{DHKS2}. Proofs were also presented for the 4m box at the N$^2$MHV level in \cite{DHKS2} and for general $k$  in \cite{QMC1}. We provide a proof of inversion symmetry below which is based on an analysis of the dual conformal properties of the $\cp_I$
factors in  the  subamplitudes. These factors are constructed from superconformal invariants, and it is clear  that the general tree
superamplitude $\ca_n^{{\rm N}^k{\rm MHV}}$ of \reef{kmhv0} is covariant
when all  momenta $p_i$ are \emph{external}.  However,  in the 1-loop formula \reef{boxcoff},  two lines of each $\ca_I$ are \emph{loop} momenta,
for example  $l_r$  and $-l_u$ for subamplitude $D$.  These lines are cut,  so that the loop momenta are null vectors. However the conformal properties
of loop line spinors $|l_i\>$ and $\eta$-variables $\eta_{l_i}$ are inherited from the one-loop environment, and these properties need to be considered carefully.

The factors $\cp_I$ are constructed from the superconformal invariants $R_{ijk}$ and the more
complicated invariants required for $k\ge 2$.  One simple but useful property of all these invariants is that they are independent of the two consecutive $\h$-variables, $\eta_i$ and $\h_{i+1}$.   At the NMHV level this  can be seen by close inspection of \reef{nmhv0}:
\begin{itemize}
  \item[i)] $\h_{i}$ does not occur in $\theta_{ki} = \sum_{q=k}^{i-1}|i\>\eta_i$\,.
  \item[ii)] $\h_{i+1}$ does not occur because the $R_{ijk}$ are defined (and appear in $\cp_I$) only for $j\ge i+2$.
\end{itemize}
For general N$^k$MHV, this property can seen from (2.14) and  (2.19)-(2.21) of \cite{DSVW}.
By cyclic symmetry of the tree superamplitudes, one can choose the `base point' $i$ to suit
one's convenience. So by choice of base point, one can arrange to make the invariants independent of \emph{any two consecutive} $\h$-variables.  Since the loop momenta always appear as \emph{consecutive} lines in the subamplitudes $A$, $B$, $C$, $D$, one can make
it manifest that the factors $\cp_I$ do not depend on the four $\h_{l_i},~ i=r,s,t,u.$  This fact implies
that the $\cp_I$ factors can be pulled out of the integral \reef{boxcoff} \emph{before} the
$\h_{l_i}$ integrals are performed, as we did in section \ref{ss:boxes} to write simple expressions for the  3m, 2mh, 2me, and 1m box coefficients.

It is also true for the 4m box, for which we write the
following expression (for N$^2$MHV
superamplitudes a different expression is given in (5.11) of \cite{DHKS2}):
 \bea
  \lab{ac4m}
  c^{\rm 4m}_{rstu} &=&\frac{1}{\Delta_{rstu}}\,
  \ca^\text{MHV}_{n,0}   \times\frac12 \sum_{\cs} \biggl[\,
   \frac{\d^{(4)}(\Sigma_{rst})\, \d^{(4)}(\Sigma_{urs})}
  {\< l_r l_s \>^4}\,\,\cyc_{\rm 4m } \cp_A \, \cp_B \, \cp_C \, \cp_D \biggr]\,,
\eea
with
\be
  \cyc_{\rm 4m} =
\frac{\cyc(1,\ldots,n)}{\cyc(A)\,\cyc(B)\,\cyc(C)\,\cyc(D)}\,,
  ~~~~~
  \Sigma_{ijk} = \< \theta_i l_i \> \< l_j l_k \>
  +\< \theta_j l_j \> \< l_k l_i \>
  +\< \theta_k l_k \> \< l_i l_j \>\,.
  \label{MIT}
\ee

To establish the dual superconformal
  properties of the $\cp_I$, we need to understand how inversion acts on them. It is sufficient to pay attention to quantities whose behavior under inversion may change because
 the $\cp_I$ occur within a  1-loop box coefficient.  Thus we need to consider the spinors $|l_i\>,~ i=r,s,t,u$ for the internal lines and the bosonic `zone vector' $x_0$ shown\footnote{The zone vector $x_0$ is not
an independent variable. Rather, it is fixed by the cut   conditions  $l_i^2 = (x_{l_i}- x_0)^2=0,~i=r,s,t,u$. }
  in Fig. 1.

To study this we first assume that subamplitude $D$ is NMHV.
(We work with $D$ because it is `massive' for all types of boxes.)
We choose base point $i=l_r$ so that  the $R_{l_rjk}$ which contribute to $\cp_D$  in \reef{nmhv0} do not depend on $\eta_{l_r}$ and $\eta_{l_u}.$  In general, a spinor for a line $|q\>$ in a tree amplitude transforms under inversion
(see \cite{DHKS1}, and also \cite{QMC1}) as
\bea \lab{invq}
\includegraphics[width=1.5cm]{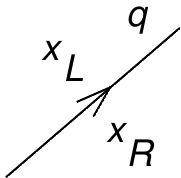}
\hspace{1.5cm}
\raisebox{6mm}{
$I \big[ |q\>_\a \big]
\,=\,  \< q|^{\b}\,
({x_{L}}^{-1})_{\b \dot{\a}}\
\,=\, \frac{1}{x_L^2}\, \< q|^{\b}\,
{x_{L}}_{\b \dot{\a}}\,
$ .}
\eea
The zone variable $x_L$ to the immediate \emph{left} (with respect to the direction) of any line determines its inversion property.\footnote{Square spinors transform with the zone variable $x_R$.} In particular, $I(|l_i\>) = \<l_i| x_i/x_i^2$ for $i=r,s,t,u$.
We distinguish between  quantities in $R_{l_rjk}$ which transform `normally,'  i.e.~in the same way they would if all lines of subamplitude $D$ were outgoing, and quantities which transform `abnormally,' because they are changed within
the 1-loop environment.
Clearly, only quantities related to the spinors $|l_r\>$ and $|l_u\>$ might transform abnormally.

Note that $l_r$ only appears in $R_{l_rjk}$ through the combination $\< l_r | x_{rj} \cdots | ~\>$. Under inversion, this transforms the `normal' covariant way,
$\< ~ | \cdots x_j^{-1} x_{jr} x_r^{-1}x_r^{-1} |l_r\> = - \< l_r | x_{rj} \cdots | ~\>/(x_r^2 \cdots)$, and the factors with $\<l_r|$ will therefore not spoil dual conformal invariance of $R_{l_rjk}$.

Since  $l_u$ comes after $l_r$ and before $u$ in the
cyclic order,  the only way it can appear in 
 $R_{l_rjk}$ 
is if
$j-1 = l_u$.
(Because  $l_u+2 \le j \le k -2$,  lines $j,k$   cannot be $l_u.$)
The spinor $|j-1\>=|l_u\>$ appears once in the numerator of
$R_{l_r  j k}$,
through $\<j-1, j\> = \<l_u u\>$,  and once in the numerator, through
$\<~| \cdots x_{kj}|j-1\>=\<~| \cdots x_{ku}|l_u\>$.  These quantities transform under inversion as
\bea
I[ \<l_u u\>] &=& - \frac{1}{x_u^2} \<l_u u\>\\
I[\<~| \cdots x_{ku}|l_u\>] &=&
\<l_u| x^{-1}_u x_u^{-1} x_{uk} x_k^{-1} \cdots |~\> = - \frac{1}{x_u^2 \cdots } \< ~ | \cdots  x_{ku}|l_u\>\,.
\eea
In both cases this is `abnormal' inversion behavior, because from the viewpoint of subamplitude $D$, we would expect $1/x_0^2$ rather than  $1/x_u^2$.
However, the abnormal factor cancels between numerator and denominator.\footnote{
In~\cite{QMC1}, the inversion factors for internal lines were taken to be arbitrary, and it was argued that the behavior of the 4m box coefficient under inversion is independent of this choice. This approach is consistent with our analysis here.}
Thus, all $R_{l_rjk}$
in $\cp_D$ transform normally; they are invariant under inversion!

The same argument applies to the more complicated invariants needed for
general N$^k$MHV subamplitudes. This can be seen by considering the
limits on the sums and the form of the invariants which occur in the definition of the general $\cp_I$; see Sec. 2 of \cite{DSVW}.  Clearly the same argument also applies when subamplitudes $A$, $B$, $C$ are  N$^k$MHV.

We now refer to \emph{any}  of the formulas for  box coefficients given in Sec. \ref{ss:boxes}, for example to
\reef{the3m}.  The box coefficient $c^{3m}_{rstu}$ is a product of four superconformal  \emph{invariant}  quantities times the MHV tree factor $\ca^{\rm MHV}_{n,0}$. We reattach the
momentum $\d$-function and write
\be  \lab{lasteq}
\d^{(4)}(\sum_{i=1}^n p_i)\ca^{\rm MHV}_{n,0}\,=\,  \d^{(4)}(\sum_{i=1}^n p_i)    \d^{(8)}(\sum_{i=1}^n |i\>\eta_i^a)/ \prod_{i=1}^n
\<i,i+1\> 
\,.
\ee
The product of $\d$-functions is invariant under inversion, so this quantity transforms \emph{covariantly} with weight factor $\prod_{i=1}^n x_i^2$.
This completes our proof that the general 1-loop box coefficients transform covariantly under inversion and are therefore also dual conformal covariant.

\setcounter{equation}{0}
\section{Discussion and Conclusions}
\lab{sec7}

It is important and somewhat subtle to distinguish between the dual conformal and dual superconformal symmetries and their action on amplitudes and superamplitudes.
\begin{enumerate}
\item The purely bosonic dual conformal symmetry has limited applicability to ordinary amplitudes, even to  the best known Parke-Taylor $n$-gluon tree approximation amplitudes.
These are constructed from spinor products $\<i \,j\>$. However, these products transform covariantly under inversion only \cite{DHKS1}  for adjacent external lines $|j\> =|i\pm1\>$.  Hence only
 the split-helicity MHV  amplitudes, $A_n( -- ++ \dots +)$, transform covariantly under dual conformal transformations.
 \item  The tree approximation \emph{super}amplitudes of $\cn =4$ SYM theory fare better. Indeed they
 are covariant under all \emph{super}conformal transformations  \cite{DHKS1,QMC1}.
 \item
Infrared divergences limit the applicability of the dual symmetries beyond the tree approximation. What is now known,
 due to the results of \cite{DHKS2} for $n=6,7,8,9$ and our work for general $n$ is that the
 ratio  ${\cal R}_{n,1}^{\rm NMHV}$ discussed in the text is dual conformal invariant. It is natural to also expect invariance under dual \emph{super}conformal transformations. However, it appears \cite{GK} that the ratios ${\cal R}_{n,1}^{\rm NMHV}$ are not invariant under all dual \emph{super}conformal transformations.
\end{enumerate}

Let us compare the structure of our final expression \reef{RwithS}   for the 1-loop  NMHV ratio
with the form conjectured in (4.62) of \cite{DHKS2}.   This form contains a sum over all
$R_{1st}$ with $4 \le s \le n-2$ and $s+2 \le t \le n$ and cyclic images and thus includes some of  the invariants which we have eliminated.  Our result shows that that form is not the `minimal' representation,  but there is no contradiction. Given our form with the dual conformal invariant $\cs$ coefficients, one can always reintroduce the $R$'s which were eliminated without
destroying the good properties of the coefficients.

The natural next step in the present program is the study of the  dual conformal properties of the ratios
${\cal R}_{n,1}^{{\rm N}^k{\rm MHV}}$ for 1-loop N$^k$MHV processes with $k>1$.  This requires
the $k>1$ superconformal invariants defined in \cite{DH} (as well as new invariants from the 4m box coefficient \cite{DHKS2}).  
They are more complicated than the $R_{rst}$ invariants we needed, and the  linear relations among them have not yet been explored. Rather than a brute force calculation, a better understanding of the structure underlying dual conformal symmetry would lead to a more natural approach to the problem.

\section*{Acknowledgments}

We are very indebted to David Kosower for suggesting this problem and early collaboration and to him and Gregory Korchemsky for much generous and  useful advice.
We also thank Fernando Alday, Nima Arkani-Hamed, Donal O'Connell, John McGreevy and Charles Thorn for useful discussions.
We are grateful for an MIT-France Seed Fund award for exchange visits with the CEA Saclay Laboratory.
HE is grateful to the Niels Bohr International Academy for hospitality during the final stage to this work.
HE is supported by NSF grant PHY-0503584.
DZF is supported by NSF grant PHY-0600465.
DZF and MK are supported by the US  Department of Energy through cooperative research agreement DE-FG0205ER41360.

\begin{appendix}
\section{The ratio $\cR_{n,1}^\text{NMHV}$ for  $n=6,7$}\label{app67}

The ratio $\cR_{n,1}^\text{NMHV}$ for $n=6,7$ requires a special treatment in our approach. A manifestly finite and dual conformal invariant form of the ratio  was previously given in~\cite{DHKS2} for $n=6,7$, and now we make contact with these results.
Recall that there is no invariant $R_{14,n-1}$ for $n=6$, and~(\ref{remNMHV}) should then be understood with ${\cal T}_{14,n-1}\to 0$. We obtain
\begin{equation}\label{6ptresult}
\begin{split}
    \cR_{6,1}^\text{NMHV}=\,&\Bigl({\cal T}_{146}+\PP^{-2}{\cal T}_{136}+\PP^{-4}{\cal T}_{135}\Bigr) R_{146} ~~+~\text{cyclic}\\
    =\,&\tfrac{1}{2}\Bigl(-2F^\M_{3456}+2F^\M_{4561}-2F^\M_{5612}-F^\MMe_{1245}- F^\MMe_{2356}-F^\MMe_{3461}+2F^\MMh_{6124}+2F^\MMh_{3451}\Bigr) R_{146} ~~+~\text{cyclic}\\
    =\,&\tfrac{1}{2}\Bigl(F^\M_{1234}-F^\M_{2345}-F^\M_{3456}+F^\M_{4561}-F^\M_{5612}-F^\M_{6123}-F^\MMe_{1245}- F^\MMe_{2356}-F^\MMe_{3461}\\
    &\qquad+2F^\MMh_{6124}+2F^\MMh_{3451}\Bigr) R_{146} ~~+~\text{cyclic}\,.
\end{split}
\end{equation}
In the last step, we have added an alternating cyclic sum of one-mass boxes integrals,
\begin{equation}
    F^{1m}_{\rm alt}=\tfrac{1}{2}\Bigl(F^\M_{1234}-F^\M_{2345}+F^\M_{3456}-F^\M_{4561}+F^\M_{5612}-F^\M_{6123}\Bigr)\,,
\end{equation}
to the coefficient of $R_{146}$. This added contribution vanishes after using cyclicity. To see this, notice that $\PP^{2k}F^{\rm 1m}_{\rm alt}=F^{\rm 1m}_{\rm alt}$ and $F^{\rm 1m}_{\rm alt}+\text{cyclic}=0$, and therefore
\begin{equation}
    F^{\rm 1m}_{\rm alt}R_{146}+\text{cyclic}\,=\,\frac{1}{3}F^{\rm 1m}_{\rm alt}(1+\PP^2+\PP^4)R_{146}+\text{cyclic}
    \,=\,\frac{1}{3}F^{\rm 1m}_{\rm alt}R_{\rm tot}+\text{cyclic}=0\,.
\end{equation}
The representation~(\ref{6ptresult}) of the $6$-point ratio coincides with the one given in~\cite{DHKS2}, and is therefore manifestly finite and dual conformal invariant.

For $n=7$,~(\ref{remNMHV3}) gives
\begin{equation}
\begin{split}\label{cR7}
    \cR_{7,1}^\text{NMHV}
    =&\,{\cal S}_{\rm tot}R_{\rm tot}+\Bigl({\cal T}_{147}+\PP^{-2}{\cal T}_{136}-(1+\PP^{-2}){\cal T}_{146}\Bigr) R_{147}\\[.5ex]
    &+\Bigl({\cal T}_{157}+\PP^{-2}{\cal T}_{137}+\PP^{-4}{\cal T}_{135}-(1+\PP^{-2}+\PP^{-4}){\cal T}_{146}\Bigr) R_{157}
    ~~+~\text{cyclic}\\
    =&\,{\cal S}_{\rm tot}R_{\rm tot}+\Bigl(-F^\M_{6712}+F^\MMh_{7124}+F^\MMh_{6714}-F^\MMM_{1246}
    -F^\MMM_{6724}\Bigr) R_{147}\\[.5ex]
    &+\Bigl(F^\M_{5671}-F^\M_{6712}-F^\M_{4567}+F^\MMe_{1245}+F^\MMh_{7125}+F^\MMh_{4561}-F^\MMM_{1246}-F^\MMM_{6724}
    -F^\MMM_{4572}\Bigr) R_{157}
    ~~+~\text{cyclic}\,,
\end{split}
\end{equation}
with
\begin{equation}
7\, {\cal S}^{(n=7)}_{\rm tot}
  =-\Li_2\big(1-  u_{1246}^{-1}
  \big)
  +\frac{1}{2}\sum_{u=4}^{5}\Li_2\big(1-u_{12u,u+1}^{-1}
  \big)
    -\ln\left(u_{7145}\right) \ln\left(u_{2367}\right)
 ~+~ {\rm cyclic}\, .
\end{equation}
The form~(\ref{cR7}) of the ratio function for $n=7$ is manifestly finite and dual conformal invariant. In fact, it is easy to see that this form agrees with the results of~\cite{DHKS2}.\footnote{In particular, the coefficients of $R_{147}$ and $R_{157}$ agree with the coefficients $V_{II}$ and $V_{I}$ of equation (4.51) in~\cite{DHKS2}.}


\section{Scalar Box Integrals}\label{appboxes}

We now present explicit forms of the dimensionless scalar box integrals $F_{rstu}$ introduced in section~\ref{secboxes}.
The scalar box integrals can be written in terms of the dual $x_{ij}$ variables. In $D=4-2\epsilon$ dimensions, one finds~\cite{Bern:1993kr}
\begin{equation}
\begin{split}
  F^{1{\rm m}}_{rstu} =&
  -{1\over\e^2} \Bigl[ (-x_{rt}^2)^{-\e} + (-x_{su}^2)^{-\e} - (-x_{ur}^2)^{-\e} \Bigr]
 \ + \Li_2\left(1-{x_{ur}^2\over x_{rt}^2}\right)
  \ + \Li_2\left(1-{x_{ur}^2\over x_{su}^2}\right)\\&
  \ +\ \frac{1}{2} \ln^2\left({x_{rt}^2\over x_{su}^2}\right)\ +\ {\pi^2\over6}
  \ + \ \Ord(\e), \nonumber \\[3mm]
  F^{2{\rm me}}_{rstu}
  =&
  -{1\over\e^2} \Bigl[ (-x_{rt}^2)^{-\e} + (-x_{su}^2 )^{-\e}
              - (-x_{st}^2)^{-\e} - (-x_{ur}^2)^{-\e} \Bigr]
  \ + \Li_2\left(1-{x_{st}^2\over x_{rt}^2}\right)
   \ + \Li_2\left(1-{x_{st}^2\over x_{su}^2 }\right) \\[1mm]
  & \ + \Li_2\left(1-{x_{ur}^2\over x_{rt}^2}\right)
   \ + \Li_2\left(1-{x_{ur}^2\over x_{su}^2 }\right) \ - \Li_2\left(1-{x_{st}^2 x_{ur}^2\over  x_{rt}^2 x_{su}^2 }\right)
   \ +\frac{1}{2} \ln^2\left({x_{rt}^2\over x_{su}^2 }\right) \ + \ \Ord(\e), \nonumber \\[3mm]
  F^{2{\rm mh}}_{rstu} =&
  -{1\over 2 \e^2} \Bigl[ (-x_{rt}^2)^{-\e} + 2(-x_{su}^2)^{-\e}
              - (- x_{tu}^2)^{-\e} - (-x_{ur}^2)^{-\e} \Bigr]
  \ + \Li_2\left(1-{ x_{tu}^2\over x_{su}^2}\right)
  \ + \Li_2\left(1-{x_{ur}^2\over x_{su}^2}\right)\\&
  \ -\frac{1}{2}  \Big(\ln\frac{ x_{tu}^2}{x_{rt}^2}  \Big)\,  \Big(\ln\frac{x_{ur}^2}{x_{rt}^2} \Big)
  \ +\frac{1}{2} \ln^2\left({x_{rt}^2\over x_{su}^2}\right)\ +\ \Ord(\e), \nonumber \\[3mm]
 F^{3{\rm m}}_{rstu} =&
  -{1\over 2\e^2} \Bigl[ (-x_{rt}^2)^{-\e} + (- x_{su}^2)^{-\e}
     - (-x_{st}^2)^{-\e} - (-x_{ur}^2 )^{-\e} \Bigr]
  \ + \Li_2\left(1-{x_{st}^2\over x_{rt}^2}\right)
   \ + \Li_2\left(1-{x_{ur}^2 \over  x_{su}^2}\right)\\[1mm]&
    - \Li_2\left(1-{x_{st}^2 x_{ur}^2 \over x_{rt}^2  x_{su}^2}\right)
     - \frac{1}{2} \Big(\ln\frac{x_{st}^2}{ x_{su}^2}  \Big)\,  \Big(\ln\frac{x_{tu}^2}{ x_{su}^2} \Big)
  -\frac{1}{2}\Big(\ln\frac{x_{tu}^2}{x_{rt}^2}  \Big)\,  \Big(\ln\frac{x_{ur}^2}{x_{rt}^2} \Big)
      \ +\frac{1}{2} \ln^2\left({x_{rt}^2 \over  x_{su}^2}\right) + \Ord(\e).  \nonumber
\end{split}
\end{equation}

\end{appendix}


\end{document}